\documentclass[10pt, conference, compsocconf]{IEEEtran}
\ifCLASSINFOpdf
\else
\fi
\usepackage{latexsym, amssymb, amsfonts, amsmath, amsthm, graphics, graphicx, subfigure, bbm, stmaryrd}

\usepackage{color}

\usepackage{graphicx}
\usepackage{verbatim}

\theoremstyle{definition}

\theoremstyle{remark}

\numberwithin{equation}{section}

\begin{document}
%
\title{Centroid estimation based on symmetric KL divergence for Multinomial text classification problem}




%
\author{\IEEEauthorblockN{Jiangning Chen\IEEEauthorrefmark{1},
Heinrich Matzinger\IEEEauthorrefmark{2},
Haoyan Zhai\IEEEauthorrefmark{3}, and
Mi Zhou\IEEEauthorrefmark{4}}
\IEEEauthorblockA{\IEEEauthorrefmark{1}
Georgia Institute of Technology,
Atlanta, Georgia 30332--0250\\ Email: jchen444@gatech.edu}
\IEEEauthorblockA{\IEEEauthorrefmark{2}
Georgia Institute of Technology,
Atlanta, Georgia 30332--0250\\ Email: matzi@@gatech.edu}
\IEEEauthorblockA{\IEEEauthorrefmark{3}
Georgia Institute of Technology,
Atlanta, Georgia 30332--0250\\ Email: hzhai8@gatech.edu}
\IEEEauthorblockA{\IEEEauthorrefmark{4}Cornell University,
Ithaca, NY 14850\\ Email: mz558@cornell.edu}}


\maketitle

\begin{abstract}
We define a new method to estimate centroid for text classification based on the symmetric KL-divergence between the distribution of words in training documents and their class centroids. Experiments on several standard data sets indicate that the new method achieves substantial
improvements over the traditional classifiers.

\end{abstract}


%
\IEEEpeerreviewmaketitle

\section{Introduction}

Text classification problem has long been an interesting research field, the aim of text classification is to develop algorithm to find the categories of given documents. Text classification has many applications in natural language processing (NLP), such as spam filtering, email routing, and sentimental analysis. Despite intensive work,remains  an  open  problem  today. 

This problem has been studied for many aspects, including: supervised classification problem, if we are given the labeled training data; unsupervised clustering problem, if we only have documents without labeling; as well as feature selection.

For supervised problem, if we assume that all the categories are independent multinomial distributions, and each document is a sample generated by that distribution, a straight forward idea is to using some linear models to distinguish them, such as support vector machine (SVM)\cite{cortes1995support,joachims1998text}, which is used to find the "maximum-margin hyperplane" that divides the documents with different labels. The algorithm is defined so that the distance between the hyperplane and the nearest sample $d_i$ from either group is maximized. The hyperplane can be written as the set of documents vector $\vec{d}$ satisfying:
\begin{equation*}
    \vec{w}\cdot\vec{d} - b = 0,
\end{equation*}
where $\vec{w}$ is the normal vector to the hyperplane. Under the same assumption, another effective classifier, using scores based on the probability of given documents conditioned on the categories, is called naive Bayesian classifier\cite{friedman1997bayesian,langley1992analysis}. This classifier learns from training data to estimate the distribution of each categories, then we can compute the conditional probability of each documents $d_i$ given the class label $C_i$ by applying Bayes rule, then the predicting of the classes is done by choosing the highest posterior probability. The algorithm to get the label for a given document $d$ is given by:
\begin{equation*}
    label(d) = \operatorname*{argmax}_j P(C_j)P(d|C_j)
\end{equation*}
When we understand the documents as sequence of words, to understand the order of the words, given the data set large enough, we can using deep learning models such as Recurrent Neural Network (RNN)\cite{tang2015document,liu2016recurrent}.

For unsupervised problem. We have traditional method SVD (Singular Value Decomposition)\cite{albright2004taming} for the dimension reduction and clustering. There also exist some algorithms based on EM algorithm, such as pLSA (Probabilistic latent semantic analysis)\cite{hofmann1999probabilistic}, which consider the probability of each co-occurrence as a mixture of conditionally independent multinomial distributions:
\begin{eqnarray*}
    P(w,d) &=& \sum_C P(C) P(d|C) P(w|C) \\
    &=& P(d) \sum_C P(C|d) P(w|C),
\end{eqnarray*}
where $w$ and $d$ are observed words and documents, and $C$ been the words' topic. As we mentioned, the parameters are learned by EM algorithm. Using the same idea, but assuming that the topic distribution has sparse Dirichlet prior, we have algorithm LDA (Latent Dirichlet allocation)\cite{blei2003latent}. The sparse Dirichlet priors encode the intuition that documents cover only a small set of topics and that topics use only a small set of words frequently. In practice, this results in a better disambiguation of words and a more precise assignment of documents to topics.

There are also many results in feature engineering, such as tf-idf\cite{ramos2003using}, n-gram, or inproved tf-idf with other feature selection\cite{schneider2004new}.

In this paper, we still assume that documents are generated according to a multinomial event model\cite{mccallum1998comparison}. We defined a new method to estimate centroid based on the symmetric KL-divergence between the distribution of documents and their class centroids, which works better than original average estimated centroid in naive Bayes method.

\textbf{Notations:} In this paper, document belong to class $j$ with index $i$ is represented as a vector $d_i^j = (x_{i_1}, x_{i_2},...,x_{i_{|V|}})$ of word counts where $V$ is the vocabulary, and each $x_{i_t} \in \{0,1,2,...\}$ indicates how often $w_t$ occurs in $d_i$. $c_i$ denotes the centroid of the class $C_i$, since we use the assumption that documents are generated according to a multinomial event model, $c_i = (c_{i_1},c_{i_2},...c_{i_{|V|}})$ satisfies: $
\sum_{j = 1}^{|V|} c_j = 1.$

\section{Our model}

Let $p = (p_1,p_2,...,p_n)$, $q = (q_1,q_2,...q_n)$ be two multinomial distributions, the KL-divergence is defined as:$$KL(p,q) = \sum_{i = 1}^n p_i\log{\frac{p_i}{q_i}}.$$

KL-divergence measures how much one probability distribution is different from another, it is strongly connected with naive bayes classifier. Given class prior probabilities $p(C_j)$ and assuming independence of the words, normalize of document vector of $d$, the most likely class for a document $d = (d_1,d_2,...,d_{|V|})$ satisfying $\sum_{i=1}^{|V|}d_i = 1$ is computed as:

\begin{eqnarray}\label{naive_bayes}
label(d) &=& \operatorname*{argmax}_j P(C_j)P(d|C_j)\\
         &=& \operatorname*{argmax}_j P(C_j)\prod_{i=1}^{|V|} (c_{j_i})^{d_i}\nonumber\\
         &=& \operatorname*{argmax}_j \log{P(C_j)} + \sum_{i=1}^{|V|} d_i\log{c_{j_i}}\nonumber\\
         &=& \operatorname*{argmax}_j \log{P(C_j)} - \sum_{i=1}^{|V|} d_i\log{\frac{d_i}{c_{j_i}}}\nonumber\\
         &=& \operatorname*{argmin}_j -\log{P(C_j)} + KL(d,c_j)\nonumber.
\end{eqnarray}

To make it symmetric of $p$ and $q$, we add in another term related to $q\log{p}$ as regularizer to get symmetric KL-divergence: $$SKL(p,q) = \sum_{i = 1}^n (p_i-q_i)\log{\frac{p_i}{q_i}}.$$ 

\begin{figure}[!hbtp]
\begin{center}
  \includegraphics[width=0.4\textwidth]{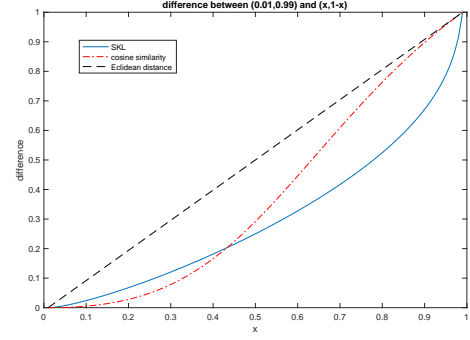}
  \caption{how difference changes between $p = (x,1-x)$ and $q = (0.01,0.99)$ in SKL, cosine similarity and Eclidean distance.}
  \label{difference}
 \end{center}
\end{figure}

To compare several measures of difference of two distributions, let $p = (x,1-x)$, $q = (0.01,0.99)$, Figure.\ref{difference} shows how the difference of two vectors change under different measures. We can see that for $p$ and $q$ far from each other, the difference of SKL decay faster, and for closer distributions, it decreases slower than linear speed. So SKL should be a good choice to distinguish distributions.

In the labeled training set, for each classes, we use SKL to find the centroid, whose sum of symmetric KL-divergence to all documents in that class reaches minimum, more specifically, the centoid is defined as following:

\begin{eqnarray}\label{mini_prob}
c_i = \operatorname*{argmin}_q \sum_{p\in C_i} SKL(p,q).
\end{eqnarray}

Let $f(q) = \sum_{p\in C_i} SKL(p,q)$, since:

\begin{eqnarray*}
f(q) &=& \sum_{j=1}^{|C_j|}\sum_{i = 1}^{|V|} (p_i^j \log{\frac{p_i^j}{q_i}} + q_i \log{\frac{q_i}{p_i^j}})\\
    &=& \sum_{j=1}^{|C_j|}\sum_{i = 1}^{|V|} p_i^j \log{p_i^j} - p_i^j \log{q_i} + q_i\log{q_i} \\
    & &- q_i \log{p_i^j}.
\end{eqnarray*}

Take partial derivative to $q_i$ we obtain: $$\frac{\partial f}{\partial q_i} = (\sum_{j=1}^{|C_j|} -\frac{p_i^j}{q_i} + \log{q_i} + 1 - \log{p_i^j}).$$

Thus:
\begin{equation*}
\left\{
\begin{aligned}
\frac{\partial^2 f}{\partial q_i^2} & =  \sum_{j=1}^{|C_j|} (\frac{p_i^j}{q_i^2}+\frac{1}{q_i})\\
\frac{\partial^2 f}{\partial q_i q_k} & =  0 
\end{aligned}
\right.
\end{equation*}

We can see that this is a convex problem. So we can obtain the global minimizer from minimization problem \ref{mini_prob}. After we get the estimation of centroid, we apply that in orginal naive bayes method \ref{naive_bayes}, under this estimator, we expected it works better than original estimator of centroid.

\section{Minimization problem}

To solve \ref{mini_prob} on the discrete probability manifold, the Wasserstein is used to get the gradient system. To this ends, suppose the graph structure $G=(V,E)$ is given where $V$ are nodes set containing all the words involved and $E$ defines the edge set which links the graph to be a connected graph. And in the examples below, the simplest histogram structure is used, that is, all the words are linked one by one in some order in a line. Also denote $n=|V|$ be the number of nodes on the graph.

Now consider a energy function $\mathcal{F}(\rho)$, let
\[
F_i(\rho)=\frac{\partial}{\partial\rho_i}\mathcal{F}(\rho)
\]
define the orientation $O$ on $G$ to be that for $(i,j)\in E$, the direction is from $i$ to $j$ if $F_i>F_j$ and that is arbitrary if $F_i=F_j$, denoting as $(i\rightarrow j)\in O$. Then the construction of the gradient of a potential function $\Phi$ based on the orientation is
\[
\nabla_G\Phi=(\Phi_i-\Phi_j)_{(i\rightarrow j)\in O},\ \ \ (\phi_i)_{i=1}^n\in\mathbb{R}^n
\]
Then, an inner product can be written as
\[
(\nabla_G\Phi,\nabla_G\Phi)_\rho=\frac{1}{2}\sum_{(i\rightarrow j)\in O}g_{ij}(\rho)(\Phi_i-\Phi_j)^2
\]
where 
\[
g_{ij}(\rho)=\left\lbrace\begin{array}{cc}
    \rho_i & \text{if }(i\rightarrow j)\in O\\
    \rho_j & \text{if }(j\rightarrow i)\in O
\end{array}\right.
\]
and the gradient flow under this metric is known as discrete 2-Wasserstein gradient flow since the discrete 2-Wasserstein distance is defined as
\begin{eqnarray*}
&&W_2(\rho^0,\rho^1)=\inf_{\rho\in\mathcal{C}}\\
&&\left\lbrace\left(\int_0^1(\nabla_G\Phi,\nabla_G\Phi)_\rho\right)^{\frac{1}{2}}:\frac{\partial\rho}{\partial t}+\nabla_G\cdot(\rho\nabla_G\Phi)=0\right\rbrace
\end{eqnarray*}
Now consider the energy function to be
\[
\mathcal{F}(\rho)=\sum_{p\in C_i}SKL(p,\rho)
\]
and the gradient flow can be written as
\[
\dot{\rho}_i+\sum_{j\in N(i)}g_{ij}(\rho)(F_i(\rho)-F_j(\rho))=0
\]

Solving this ODE obtains the solution for problem \ref{mini_prob}.

\section{Experiment}

\begin{figure}[!hbtp]
\begin{center}
  \includegraphics[width=0.4\textwidth]{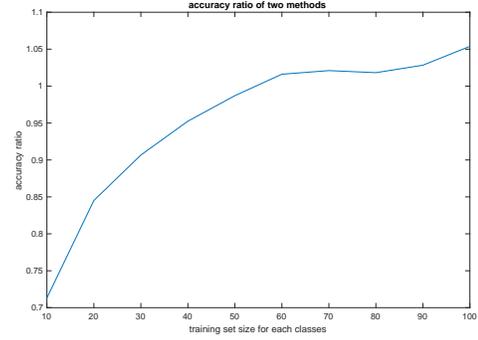}
  \caption{Average accuracy ratio under seven topics.}
  \label{accuracy}
 \end{center}
\end{figure}

We applied our method on seven topics of single labeled documents in Reuters-21578, we find the accuracy of naive bayes using our centroid estimator increasing faster than original method, see Figure.\ref{accuracy}, and when training size is large enough, our method achieves substantial improvements over the traditional method.

\begin{figure}[!hbtp]
\begin{center}
  \includegraphics[width=0.4\textwidth]{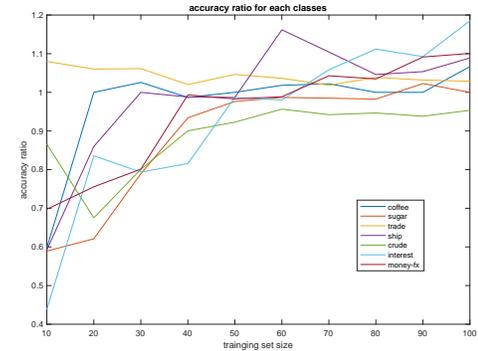}
  \caption{Accuracy ratio for seven topics.}
  \label{accuracy_class}
 \end{center}
\end{figure}

For each single class, the behave of our method versus traditional naive bayes estimator can be find in Figure.\ref{accuracy_class}. We can a clear increasing trend for topics as training size becoming larger. 

\begin{table}
\setlength{\tabcolsep}{2mm}{
\begin{tabular}{|c|c|c|c|c|c|c|}
\hline
coffee& sugar& trade& ship& crude& interst& money-fx\\
\hline
 9.0348& 8.9305&    6.2703&    9.1293&    7.3662&    7.4778&    6.9361\\
\hline
\end{tabular}}
\caption{average SKL to other classes}
\label{average_SKL}
\end{table}

Table.\ref{average_SKL} shows the average SKL to other classes, from Figure.\ref{accuracy_class} we can see that class 'trade' is the only one doesn't have trend of increasing, that might because it is very closed to other classes, and SKL cannot distinguish it well based on our observation in Figure\ref{difference}.

\section{Open problems}

\begin{itemize}

    \item We find better estimator for centroid using naive bayes, can we find similar result for other estimators?
    \item Can this centroid estimator be extended to be used in unsupervised learning problem?
    \item When we solve the minimization problem, we have a graph structure for each feature. We are using a connecting graph now, can we use the partially connected graph to demonstrate correlation of words?
    
\end{itemize}

\end{document}